\documentclass[fleqn,10pt]{wlscirep}
\usepackage[utf8]{inputenc}
\usepackage[T1]{fontenc}
\title{A quasi-monoenergetic short time duration compact proton source for probing high energy density states of matter.}

\author[1]{J. I. Apiñaniz}
\author[1]{S. Malko}
\author[2]{R. Fedosejevs}
\author[4]{W. Cayzac}
\author[4]{X. Vaisseau}
\author[1]{D. de Luis}
\author[1]{G. Gatti}
\author[6]{C. McGuffey}
\author[6]{M. Bailly-Grandvaux}
\author[6]{K. Bhutwala}
\author[3,4,5]{V. Ospina-Bohorquez}
\author[3]{J. Balboa}
\author[5]{J.J. Santos}
\author[5]{D. Batani}
\author[6]{F. Beg}
\author[1]{L. Roso}
\author[1]{J. A. Perez-Hernandez}
\author[1,7,8]{L. Volpe}

\affil[1]{Centro de Laseres Pulsados (CLPU), Parque Cientifico, E-37185 Villamayor, Salamanca, Spain}
\affil[2]{University of Alberta, Department of Electrical and Computing Engineering. Edmonton, Alberta, Canada T6G 2V4}
\affil[3]{University of Salamanca, Salamanca, Spain}
\affil[4]{CEA, DAM, DIF, F-91297 Arpajon, France}
\affil[5]{University of Bordeaux, CNRS, CEA, CELIA (Centre Lasers Intenses et Applications), UMR 5107, F-33405 Talence, France}
\affil[6]{ CMEC (Center of Matter under Extreme Conditions), University of California San Diego, La Jolla CA 92093, United States}
\affil[7]{Laser-Plasma Chair at the University of Salamanca, Salamanca, Spain}
\affil[8]{Instituto Universitario Física Fundamental y Matemáticas, 37008-Salamanca, Spain}
\affil[*]{japinaniz@clpu.es}

\begin{abstract}
We report on the development of a highly directional, narrow energy band, short time duration proton beam operating at high repetition rate, suitable for measurements of stopping power in high energy density plasmas as well as other applications. The protons are generated with an ultrashort-pulse laser interacting with a solid target and converted to a pencil-like narrow-band beam using a compact magnet-based energy selector. We experimentally demonstrate the production of a proton beam with an energy of 500 keV and energy spread well below  $ 10 \% $, and a pulse duration of 260 ps. The energy loss of this beam is measured in a 2 $\mu$m thick solid Mylar target and found to be in within 1$\%$ of theoretical predictions. The short time duration of the proton pulse makes it particularly well suited for applications involving the probing of highly transient plasma states produced in laser-matter interaction experiments, in particular measurements of proton stopping power.

\end{abstract}
\begin{document}

\flushbottom
\maketitle
%
%
\thispagestyle{empty}

\section*{Introduction}
During the last decades, multi-Terawatt and Petawatt laser facilities have become standard across the world \cite{Neely:2015}, allowing the study of new regimes of laser-plasma interaction. Short pulse lasers with intensities above 10$^{18}$ W/cm$^2$ have been widely used for the generation of compact, high brightness particle and radiation sources. One of the most commonly used techniques for proton source generation is the mechanism known as Target Normal Sheath Acceleration (TNSA) \cite{Wilks:2001,Macchi:2013}. It opens up the possibility to generate multi-MeV protons up to 100 MeV energy  \cite{Higginson:2018} and ion beams with high brilliance and relatively short time duration at high repetition rate. The precise energy deposition of protons in matter, due to the existence of the Bragg Peak, is an important characteristic for many applications, including proton therapy \cite{Ken:2014, Malka:2019}, ion-induced isochoric heating of matter \cite{Patel:2013,McGuffey:2020, Pelka:2010,Bhutwala:2020}, proton radiography \cite{Mackinnon:2006,Volpe:2011}, alpha particle heating in inertial confinement fusion (ICF) \cite{Hurricane:2016, Zylastra:2019}, the proton fast ignition approach to ICF \cite{Roth:2001}, and heavy ion fusion \cite{Fernandez:2014, Hoffman:2018}. 

The energy spectrum of laser-driven proton beams, however, is typical very broad (tens of MeV), whilst many applications require a narrow energy band proton beam or at least a precise control of the selected energy range. In these cases the generation of proton beams via laser-matter interaction needs to be complemented with a secondary system for the energy selection and beam transport of such proton sources.
Several methods have been developed to select specific ion energies and to transport the beam \cite{Toncian:2006,Chen:2014, Teng:2016, Jahn:2019,Brack:2020}. Here we report a novel design for such a proton energy selector and its experimental characterization performed at the Centro de Laseres Pulsados (CLPU) VEGA 2 laser facility \cite{Volpe:2019}. The system allows one to select a bunch of laser generated protons with an energy spread of few percent (in the presented case a proton beam of 498 $\pm$ 4 keV with a minimum energy bandwidth $\Delta$E = 33 $\pm$ 4 keV, $\Delta$E/E = 6 $\pm$ 0.75 $\%$). It is based on the use of a 1.2 T compact permanent dipole magnet, with a longitudinal dimension of $\sim$ 6 cm. The geometry allows the very close location of the selector to the proton source. The main features of this energy selector are the small energy bandwidth and the short time spread of the selected proton beam. These are both necessary properties for many applications where the probed target evolves in time or where the characteristic time of interaction is short and below the ns time scale. A minimum time spread of 260 $\pm$ 15 ps is obtained in the setup described here. One of the applications demanding a short time spread of the proton beam is proton stopping power measurements in extreme states of matter \cite{Cayzac:2017, Frenje:2019, Zylstra:2015, Malko:2020}. Such states of matter exist for a short time and their parameters change on a sub-nanosecond time scale.

\section*{Results}
\subsection*{Energy selector optimization}
\begin{figure}[ht]
\centering
\includegraphics[width=0.7\linewidth]{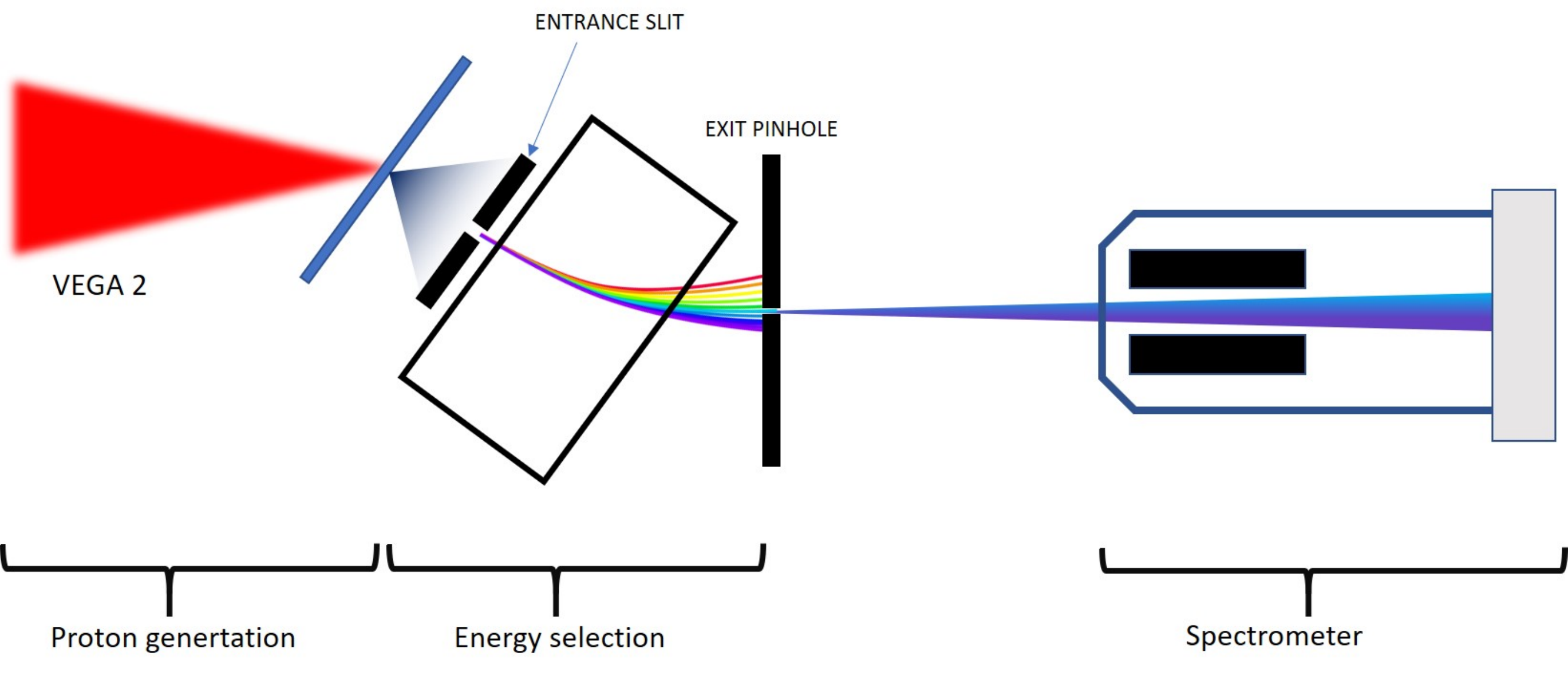}
\caption{Scheme of the experimental setup in three stages: proton generation, energy selection using magnet based proton selector and proton energy measurement with the magnet spectrometer.}
\label{fig:Setup}
\end{figure}
The characterization of the proton energy selector was performed at the CLPU using the 30 fs, 200 TW system VEGA 2. A scheme of the experimental setup is shown in Figure \ref{fig:Setup}. 
The experimental scheme consists of three main stages: (i) the VEGA 2 laser is used to accelerate a broad-spectrum proton through the TNSA mechanism, (ii) the energy selector selects out a proton bunch at a specific energy with narrow energy and angular spread, and (iii) the selected proton bunch is measured with a magnet-based spectrometer. A central proton energy of 500 keV was chosen for this setup, to be used in a subsequent measurement of ion stopping power in matter [Malko2020 submitted for publication]. 
The entrance slit and exit pinhole of the selector are adjusted to optimise the proton flux and energy bandwidth at the desired energy. The high repetition rate of the system allowed a statistical characterization of the proton energy selection. 

\begin{figure}[ht]
\centering
\includegraphics[width=0.55\linewidth]{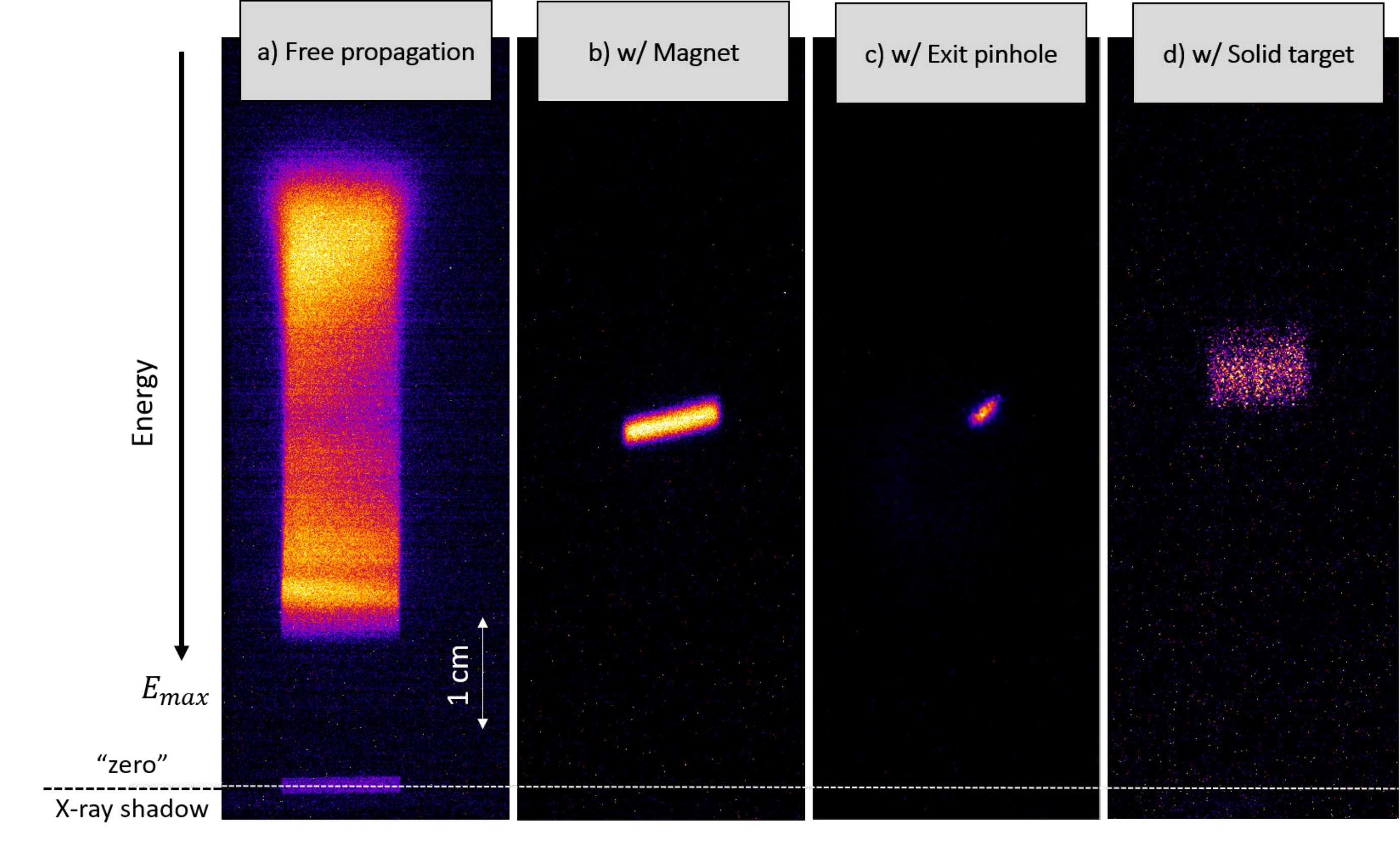}
\caption{MCP traces in three configurations: a) Full proton beam with no magnetic selector in (full TNSA spectrum). b) Partial selection magnetic selector in with only the 20 $\mu$m entrance slit. c) Full selection with 20 $\mu$m entrance slit and  20 $\mu$m exit pinhole). d) Selected beam c) after passing through a solid 2 $\mu$m  Mylar foil. }
\label{fig:Exp_MCP}
\end{figure}

\begin{figure}[ht]
\centering
\includegraphics[width=0.75\linewidth]{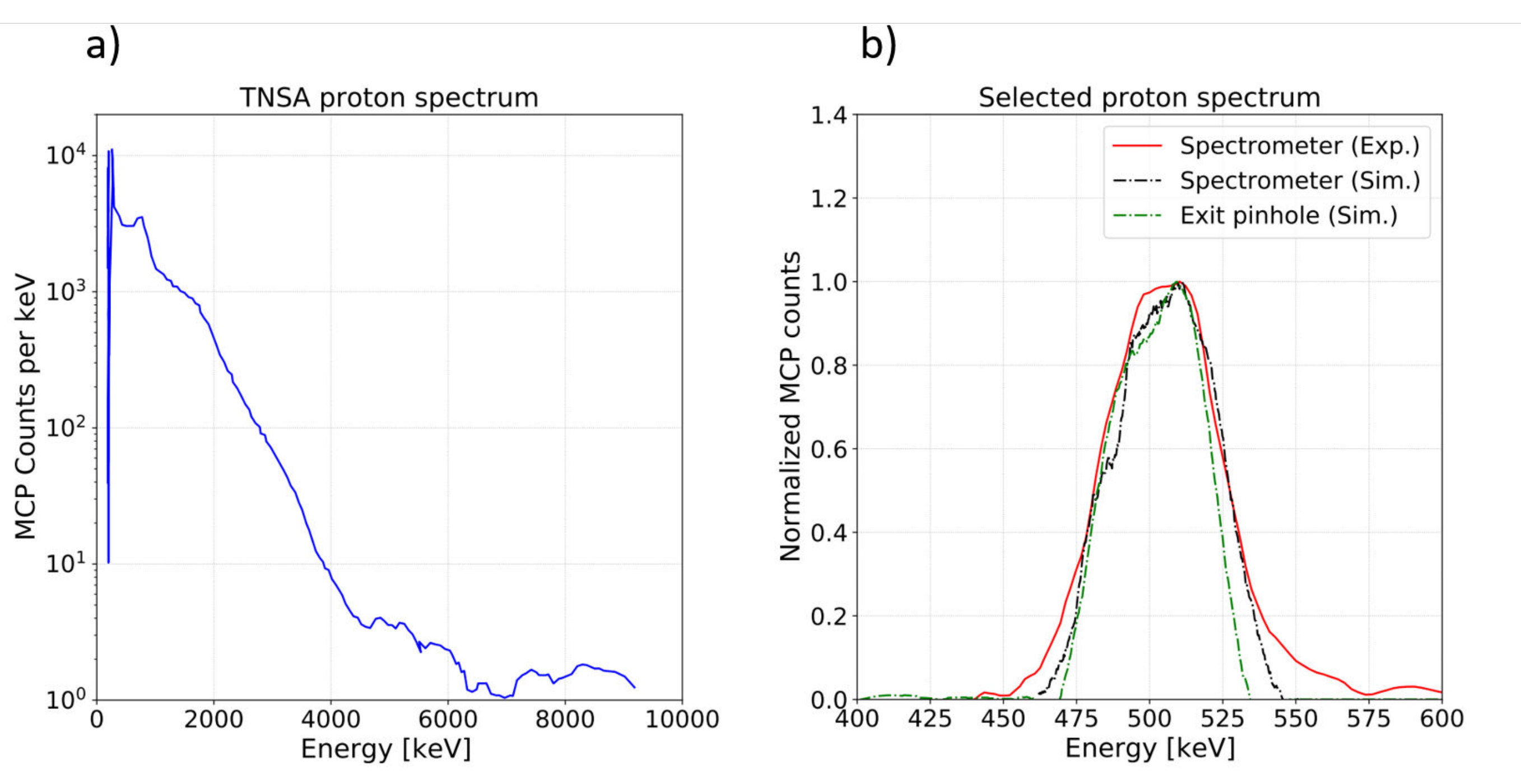}
\caption{ Experimental spectra measured with magnet spectrometer. a) Spectrum obtained in configuration of no selection featuring initial proton spectrum with cut-off energy of 4 MeV. b) The selected 500 keV proton beam spectrum obtained in full selection using 20 $\mu$m entrance slit and 20 $\mu$m exit pinhole (red solid curve), synthetic proton spectrum obtained with MC FLUKA simulations at the exit pinhole (green dashed curve) and on spectrometer (black dashed curve) in corresponding experimental geometry. }
\label{fig:Exp_MCPSpectrum}
\end{figure}

Figure \ref{fig:Exp_MCP} a) shows the original TNSA proton beam as measured without the selector. 
In Figure \ref{fig:Exp_MCP} b) the magnet selector was placed in and only the protons passing through the first entrance slit (20 $\mu$m aperture) were entering the spectrometer. Figure \ref{fig:Exp_MCP} c) refers to the final proton energy selection with placing the exit pinhole (20 $\mu$m diameter size) after the magnet. The corresponding energy spectra for the full TNSA proton beam [Fig.\ref{fig:Exp_MCP} a)] and selected proton beam [Fig.\ref{fig:Exp_MCP} c)] are shown in Figure \ref{fig:Exp_MCPSpectrum}. 
The proton beam profile at the MCP detector is the result of dispersion of the beam in the horizontal plane resulting from the magnetic selector (magnetic field oriented in vertical direction) with an additional apparent rotation due to the energy dependent deflection of the protons entering the spectrometer. Finally, figure \ref{fig:Exp_MCP} d) shows the pencil-like proton beam after was passing through the 2 $\mu$m  Mylar solid foil that will be discussed in detail in the following sections.

The selected proton beam was characterized with various apertures for the slit and pinhole starting with a 50 $\mu$m entrance slit and a 200 $\mu$m exit pinhole, down to both a slit and pinhole of 20 $\mu$m in size. The energy bandwidth (FWHM) obtained for each of configuration is shown in the Table \ref{tab:example}.

\begin{table}[ht]
\centering
\begin{tabular}{|l|l|l|l|}
\hline
Entrance slit & Exit pinhole/slit & Central energy [keV] & Bandwidth [keV]\\
\hline
 50 $\mu$m slit & 200 $\mu$m slit  & 501 $\pm$ 8 & 79 $\pm$ 8 \\
\hline
 50 $\mu$m slit & 20 $\mu$m pinhole & 510 $\pm$ 4 & 54 $\pm$ 4 \\
\hline
 20 $\mu$m slit & 20 $\mu$m pinhole & 498 $\pm$ 4 & 44 $\pm$ 4 \\
\hline
 20 $\mu$m slit & 10 $\mu$m pinhole  & 497 $\pm$ 4 & 33 $\pm$ 4 \\
\hline
\end{tabular}
\caption{\label{tab:example} The energy bandwidth of selected $\sim$ 500 keV proton beam measured for each of energy selector configurations of entrance slit and exit pinhole }
\end{table}

The values of the bandwidth presented in Table \ref{tab:example} represent the average values of the energy bandwidth over N shots with a total error estimated as $\sigma_{tot} =\sqrt{\sigma _{stat}^{2}+ \sigma _{sys}^2}$, where the statistical error is $\sigma _{stat} = \sigma/\sqrt{N}$ and  $\sigma_{sys}\approx \pm$ 2.5 keV is the systematic error coming from the uncertainty in the vertical position of the measured proton signal with respect to the zero-deflection point at the MCP detector. 
The selected proton beam spectrum shown in Fig. \ref{fig:Exp_MCPSpectrum} b) features a central energy is $E_{c}$= 498 $\pm$ 4 keV with a energy bandwidth of $\Delta_{E_c}$ = 44 $\pm$ 4 keV.  Such error on the central energy and energy bandwidth suggests that the selected proton energy and energy bandwidth has low sensitivity to the laser shot-to-shot instability (pointing stability $\sim$ 12 $\mu$m, energy variation $\sim$ 3 $\%$). 
The configuration using the 20 $\mu$m entrance slit and 20 $\mu$m has an optimal performance in terms of both a small energy bandwidth and a reasonable proton flux ($\sim$ 1500 detected protons at the MCP). The absolute number of protons is estimated using an MCP calibration made with proton energies < 1 MeV \cite{Prasad:2013}. 
The selected proton beam has been also characterized in terms of spatial size for this configuration. The proton beam spot of 50 $\mu$m diameter has been measured 0.9 cm away from the exit pinhole using Radiochromic Film (RCF) placed on the propagation proton propagation axis after selection. The result well agrees with the observed proton signal size at MCP spectrometer $\sim$ 1 mm (60 cm from the exit of pinhole).

\subsection*{Energy loss measurement in solid target}

We used this setup to measure the energy loss of a 510 keV proton beam in a solid cold target of 2 $\mu$m Mylar coated with 40 nm Aluminum, having a total areal density of 2.4 $\pm$ 10 $\%$ g/cm$^{2}$. The target has been placed on the proton beam axis 0.9 cm away from the exit pinhole of selector. The experimental proton signal observed on the MCP detector after passing through the cold sample is shown on the Figure \ref{fig:Exp_MCP} d). One can see that the observed proton signal on the MCP has been modified in shape and shifted up towards lower energy. The modification of the proton beam profile is directly related to the proton beam straggling and scattering in the solid target \cite{Rossi:1941, Volpe:2011}. The proton beam scattering complicates the estimation of the central energy of the proton beam at the MCP detector and thus increases error of the energy loss measurement. However the insertion of an additional horizontal slit in front of the spectrometer reduces the scattering effect on the MCP detector and allows for a more precise measurement of the central energy. The multiple scattering of protons in matter can be considered by estimating the mean scattering angle $\theta$ = $\sqrt{A[\text{g/cm}^{3}]}/E [\text{MeV}]$ \cite{Volpe:2011} for a proton pencil-like beam of 510 keV entering a 2.4 g/cm$^{2}$ areal density target which gives $\theta \sim $3$^\circ$. Such a scattering angle produces a beam much larger than the entrance of the spectrometer. The FLUKA simulations find that a partial collection of the protons increases the measurement uncertainty in the central energy by up to 3.5 keV.  

\begin{figure}[ht]
\centering
\includegraphics[width=0.45\linewidth]{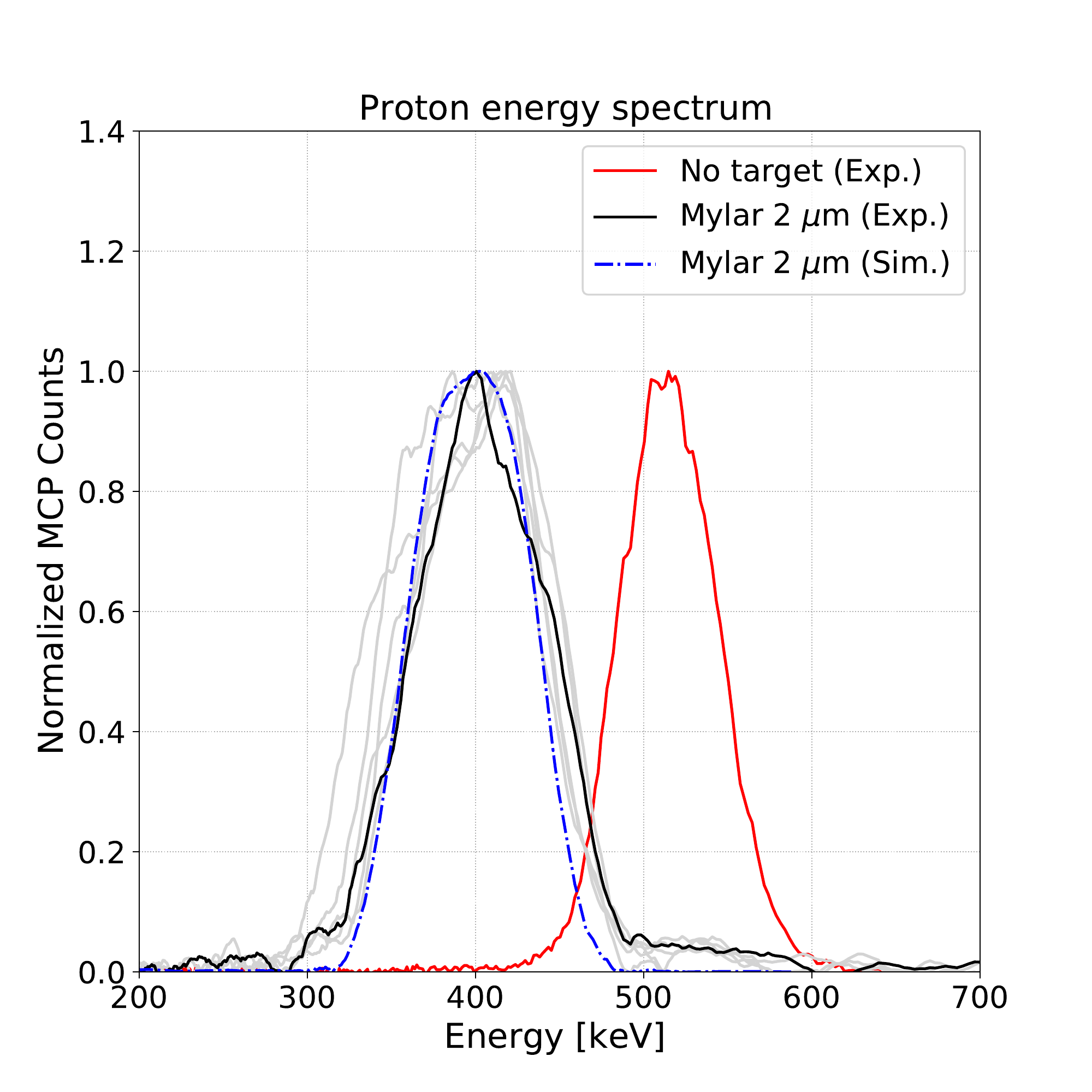}

\caption{The experimental selected proton spectrum (red curve) and the downshifted proton spectrum after passing through 2 $\mu$m Mylar coated with 40 nm Aluminum (black curve). The additional spectra  (grey curves) are shown to demonstrate the repetitive behaviour. The simulated downshifted spectrum (dashed blue curve) is obtained with FLUKA MC simulation. }
\label{fig:Solid target}
\end{figure}

Figure \ref{fig:Solid target} demonstrates the optimized initial proton spectrum after the exit pinhole with central energy of 510 keV and energy spectra after passing through the solid target. The downshifted experimental spectra is also compared with simulated spectrum in solid target at the spectrometer with FLUKA MC simulations. 
The measured energy loss of the 510 keV proton in cold target is 106 $\pm$ 7 keV. The total error consists of $\sqrt{\sigma ^2_{stat}+\sigma^2_{sys}}$, where $\sigma_{stat} = \pm$ 6 keV and systematic error of $\sigma_{sys} = \pm$ 3.5 keV coming from the aforementioned partial collection of protons. The measurement agrees with the theoretical prediction using SRIM \cite{SRIM} of 107 keV. 
The results of the energy loss is very repetitive and in good agreement with the theoretical predictions while the discrepancy of $\pm$ 7 keV is well within the $10\%$ uncertainty in the sample thickness as specified by the manufacturer equivalent to energy loss error of $\pm$ 11 keV. It indicates that the real sample thickness is quite close to its nominal value.

\subsection*{Simulations}

\begin{figure}[ht!]
\centering
\includegraphics[width=0.9\linewidth]{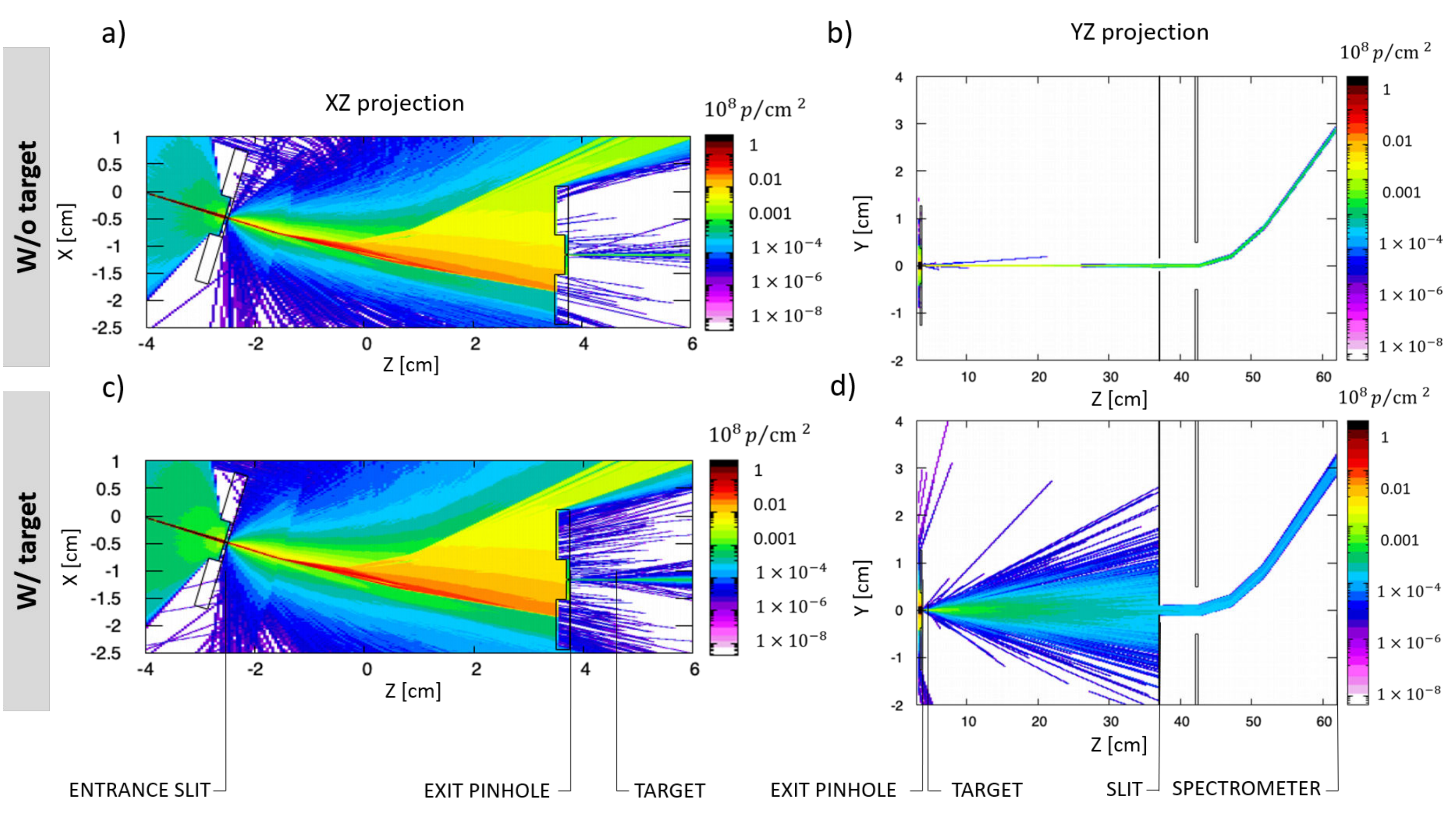}
\caption{Monte-Carlo simulations performed with the FLUKA MC code.  a) The protons enter through the entrance slit, are deflected by the $B_{y} = 1.2$ T magnetic field, and 500 keV protons are selected by the exit pinhole. b) Propagation of protons and deflection by the $B_{x}=0.2$ T magnetic field of MCP spectrometer. c) and d) Equivalent simulations including a 2 $\mu$m Mylar target. Scattering in the target increases the proton divergence and only part of the beam is sampled by the entrance pinhole of the spectrometer.}
\label{fig:FLUKA}
\end{figure}

The experimental results were modeled with 3D Monte-Carlo simulations using the FLUKA code \cite{Bohlen:2014,Ferrari:2005}.
Figure \ref{fig:FLUKA} a) shows the proton fluence in the energy selector for the experimental configuration of a 20 $\mu$m  entrance slit and 20 $\mu$m exit pinhole. Figure \ref{fig:FLUKA} b) presents the propagation of the selected 500 keV proton beam from the pinhole to the MCP screen, where the proton beam is deflected vertically by the magnetic field of the spectrometer.

A synthetic spectrum at the exit of the pinhole was acquired using an absolute energy detector with a resolution of 2 $\times$ $10^{-4}$ MeV integrated over the solid angle as shown in Figure \ref{fig:Exp_MCPSpectrum} b). A synthetic image of the proton signal on the MCP screen with a resolution of 7 $\mu$m [Fig. \ref{fig:Dis_MCP} b)] was used for comparison with the experimental image on MCP. 

Figure \ref{fig:Exp_MCPSpectrum} (b) shows a comparison between the simulated spectra obtained at the exit of the pinhole by the absolute detector (dashed green curve) and by the spectrometer (dashed black curve) considering a source size of 150 $\mu$m and the experimental spectrum on the MCP (solid red curve). The simulation results show a modification in the final spectra clearly induced by the proton divergence that can be seen by comparing the spectrum at the exit of the pinhole and at spectrometer. However this modification only affects the lower part of the distribution modifying its wings. This indicates that the measurement of energy bandwidth (FWHM) at the exit of the pinhole $\Delta E_{P}$ is in agreement with the energy bandwidth on the spectrometer $\Delta E_{S}$ within 3 keV.
A parametric study of the spatial size and the energy bandwidth of the selected proton beam as a function of the initial real proton source size has been performed in order to find the best fit to the typical experimental signal with an energy bandwidth of 44 keV and proton trace thickness of $\sim$ 1 mm. The results of simulations shown in Table \ref{tab:FLUKA table} suggest that one would need to have a proton source size of 150 $\mu$m to match the features of the proton beam at the MCP. We finally conclude that the proton beam divergence has a small effect on the final bandwidth measured by the spectrometer and this effect can be easily mitigated by moving the magnet spectrometer closer to the exit pinhole. 
In order to calculate the time spread of the corresponding selected proton beam in FLUKA, an artificial time of flight calculation was implemented, assuming  $t_{0}$ = 0 at the position of the proton source. It was found that the selected beam of 500 keV central energy and 44 keV $\pm$ 4 bandwidth leaves the exit pinhole at 8.8 ns with a time spread of 360 $\pm$ 15 ps (FWHM), which is also in agreement with trajectory simulations.

The simulations of proton beam propagation after passing though solid target has been performed in order to study the scattering effect and its mitigation for the proton stopping power measurements. The Figure \ref{fig:FLUKA} c) and d) shows the equivalent simulation of proton energy selection Figure \ref{fig:FLUKA} a), b) including 2 $\mu$m Mylar coated with 40 nm Al target. The target placed 0.9 cm away from the exit pinhole produces scattering effect that dominates over the initial proton divergence. The scattered beam is artificially reduced by the 2 mm horizontal slit located in front of the 1 cm entrance of the spectrometer as shown in \ref{fig:FLUKA} d). The slit decreases the number of entering particles drastically, however it reduces the size the of scattered beam on MCP allowing precise estimation of the peak energy. Such sampling of the proton beam yields in the error of 3.5 keV calculated by the comparison absolute downshifted spectra obtained with full and partial collection of the protons at the entrance of the spectrometer.

\section*{Discussion}
In this section we discuss the effect of the proton divergence in the analysis and performance of the adjustable proton energy selector. 

\subsection*{Estimation of initial proton source size}
It is well known that laser produced proton sources can exhibit sizes much bigger than the laser spot \cite{Borghesi:2004,Borghesi:2006}. The source size can range to several hundreds of microns for the lowest proton energies. Under these conditions the proton divergence is mainly controlled by the source size and the distances to the slit and the pinhole. The divergence of the proton beam plays a double role: on the one hand, it creates a geometrical magnification of the beam at the MCP detector. On the other hand it is responsible for the minimum achievable bandwidth i.e. the more divergence, the more spread in the energy of the trajectories that can pass through the selector.

\begin{figure}[ht]
\centering
\includegraphics[width=0.6\linewidth]{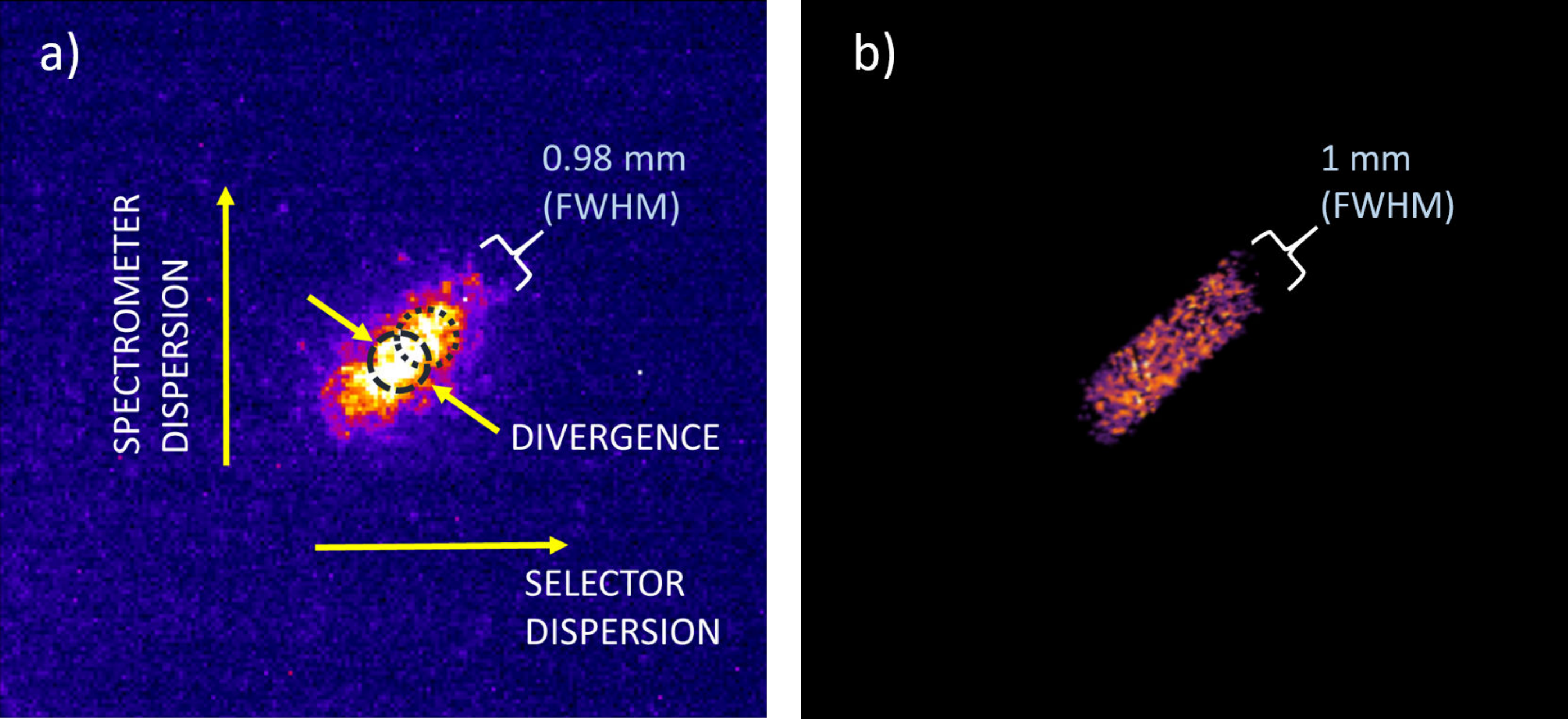}
\caption{a) Typical proton signal on MCP obtained with configuration of 20 $\mu$m slit and 20 $\mu$m pinhole and featuring 44 keV energy bandwidth (FWHM) and proton signal width of $\sim$ 0.98 mm. b) Proton signal on MCP obtained with FLUKA simulation reproducing an experimental result of energy bandwidth of 44 keV (FWHM) and width of $\sim$ 1 mm using initial source size of 150 $\mu$m. }
\label{fig:Dis_MCP}
\end{figure}

An example of the spectrum obtained with the selector configuration of 20 $\mu$m entrance slit and 20 $\mu$m exit pinhole is shown in Figure \ref{fig:Dis_MCP} a).
The trace is the typical result of the magnetic dispersion from the selector (horizontal dispersion) and the spectrometer (vertical dispersion). These combination leads to the aspect ratio of the trace. The width of the trace gives a direct measurement of the geometrical divergence. A simple pinhole magnification model can be used to deduce the initial proton source size from the width of the MCP proton trace. The proton source is imaged onto the MCP screen after passing through an ideal point-like pinhole. The magnification of the source is estimated by using the distance from the proton source to the pinhole (7.8 cm) and the distance from  the pinhole to the MCP detector (60 cm) and by using the measured trace width on the detector which is $\sim$ 0.98  mm (FWHM) as shown in Figure \ref{fig:Dis_MCP} a). The final magnification is calculated to be 7.7, which corresponds to an initial proton source size of around 150 $\mu$m, which is in agreement with a more detailed calculation performed using the Monte Carlo code. 

\subsection*{Modelling of the minimum selected bandwidth}
\begin{figure}[ht]
\centering
\includegraphics[width=0.7\linewidth]{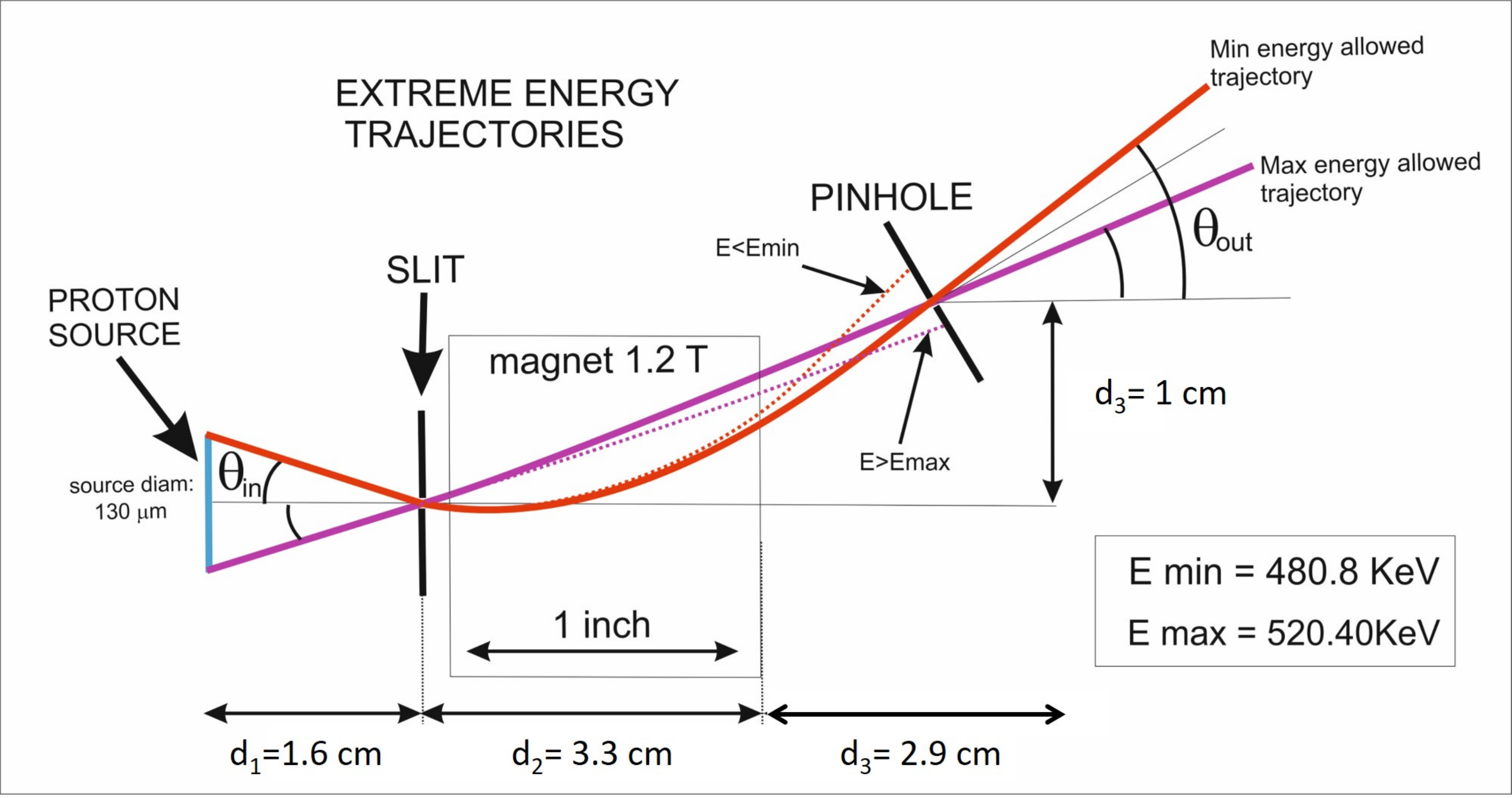}
\caption{Scheme of maximum and minimum allowed energies due to horizontal divergence. Estimated values using simple gyroradius model in a square B field are given.}
\label{fig:Dis_SM}
\end{figure}

The effect of the divergence in the bandwidth can be modeled  by using simple trigonometry as shown in Figure \ref{fig:Dis_SM}. This simple model shows the effect of the source size on the energy bandwidth at the exit pinhole ($\Delta E_{P}$). Minimum and maximum energy allowed trajectories are graphically shown and estimated as a function of incidence angle $\theta_{in}$ and energy using a simple 2D model of proton gyro-radius in a flat-top B-Field. 

Monte Carlo simulations were employed to study the effect of the proton source size on the energy bandwidth at the exit pinhole $\Delta E_{P}$ and at the spectrometer $\Delta E_{S}$ that corresponds to the experimental measurements shown in Table \ref{tab:FLUKA table}. The simulation results show that there is no major effect of divergence on the measurement of the energy bandwidth at the spectrometer ($\Delta E_{P} \approx$ $\Delta E_{S}$).
The example of the MCP trace shown in Figure \ref{fig:Dis_MCP} a) has been reproduced with FLUKA simulations. The simulations required a proton source size of 150 $\mu$m to obtain the proton signal of 1 mm width [Fig. \ref{fig:Dis_MCP} b)] which is in close agreement with the observed experimental proton signal [Fig. \ref{fig:Dis_MCP} a)].
\begin{table}[ht]
\centering
\begin{tabular}{|l|l|l|l|}
\hline
Source size [$\mu$m] & Width [mm] & $\Delta E_{P}$ [keV]& $\Delta E_{S}$ [keV] \\
\hline
100&  0.72& 27.5& 33.8 \\
\hline
110& 0.77& 32.1& 32.5 \\
\hline
120& 0.88& 34.4& 35\\
\hline
130& 0.96& 36.9& 39 \\
\hline
\textbf{150}& \textbf{1.00} & \textbf{41.9} & \textbf {43.6} \\
\hline
180& 1.19& 51.2& 54\\
\hline
\end{tabular}
\caption{\label{tab:FLUKA table} Selected 500 keV proton beam bandwidth and spatial size as a function of the proton source size obtained with FLUKA MC simulations. The best fit to the experimental result is highlighted. }
\end{table}
Moreover, it is in a good agreement with a source size of $150$ $\mu$m previously estimated by pinhole imaging magnification.
The simulations also agree with the simple model prediction of energy bandwidth $\Delta E_{P}$ scaling as a function of the proton source size.

The simulation results indicate that in order to further decrease the energy bandwidth one would need to reduce the experimental proton source size using additional techniques such as wire-targets \cite{Covan:2004}, self-generated magnetic fields \cite{Malko:2019} or mass-limited targets \cite{Buffechoux:2010} or by introducing an additional pinhole in the magnet selector with a consequent reduction of the number of selected protons. 

When the proton beam is used for proton stopping power measurement the final output is completely dominated by the proton multiple scattering in the sample, magnifying the final beam size. Despite the increase of scattering the final beam shows a high degree of spatial and energy uniformity and this permits a measurement in the reduced area of the beam without losing energy precision. In particular, it is possible to use a thin slit (along the Lorenz force direction) to increase the final spectral resolution.     

In summary, we present a new adjustable platform for proton energy selection from laser-plasma sources that has been designed, characterized and experimentally tested for measuring proton stopping power in a solid plastic target. The novel design produces a narrow-band proton beam with a short time spread, as is required for numerous applications, and in particular, proton stopping power measurements in transient sates of matter. After optimization the selection of proton projectiles of 498 $\pm$ 4 keV central energy with an energy bandwidth of 44 $\pm$ 4 keV and a time spread of 360 $\pm$ 15 ps at the exit pinhole with the required signal level has been demonstrated using 20 $\mu$m slit and 20 $\mu$m pinhole apertures in the energy selector.
Different configurations of the proton energy selector have been experimentally characterized in terms of the proton signal intensity and energy bandwidth. The energy bandwidth was gradually reduced from 79 keV to 33 keV by using the pinholes of smaller sizes from 200 $\mu$m to 10 $\mu$m with a time spread ranging from 630 ps to 260 ps. Numerical simulations performed with FLUKA have been used to interpret the experimental results and demonstrate the effect of the proton source size on the intrinsic bandwidth of the selected protons and on the shape of the signal on the MCP. 
As a demonstration of feasibility of the platform, we also show the energy loss measurement of the selected proton beam in a 2 $\mu$m thick plastic foil that is in good agreement with theoretical predictions. It is expected that the current platform can serve as a short pulse tunable source of protons for a variety of applications in the future.

\section*{Methods}

\subsection*{Experimental setup}

\noindent The basic characteristics of VEGA II and its main beam transport system have been explained in more detail elsewhere\cite{Volpe:2019}. 
In order to accelerate protons via the TNSA mechanism the VEGA 2 laser pulse is transported and focused by an F/13 (F = 130 cm)  parabolic mirror onto a 3 $\mu$m thick aluminium foil at a 14.5$^\circ$ incidence angle within a 20 $\mu$m focal spot (FWHM) yielding an intensity on target of $\sim 10^{19}$ $\text{W}/\text{cm}^{2}$. To provide fast target replacement after each laser shot, a motorized sandwich target holder with 45 $\times$ 45 800 $\mu$m apertures was used.  
The second stage of the proton energy selection is based on energy dispersion by a 1.2 T permanent magnet with adjustable entrance and exit apertures to control the energy bandwidth of the proton beam. 
The selected beam characterization is done by means of a magnetic spectrometer (0.2 T dipole magnet) coupled with a phosphor MCP detector and an imaging system, indicated as a third stage.  Selected beam spectra were recorded at high repetition rate (0.1 Hz) with an energy resolution of 2 keV per pixel at 500 keV. Energy selection and measurement are explained in detail in the following sub-sections. 

\subsection*{Energy selector}
\begin{figure}[ht]
\centering
\includegraphics[width=0.9\linewidth]{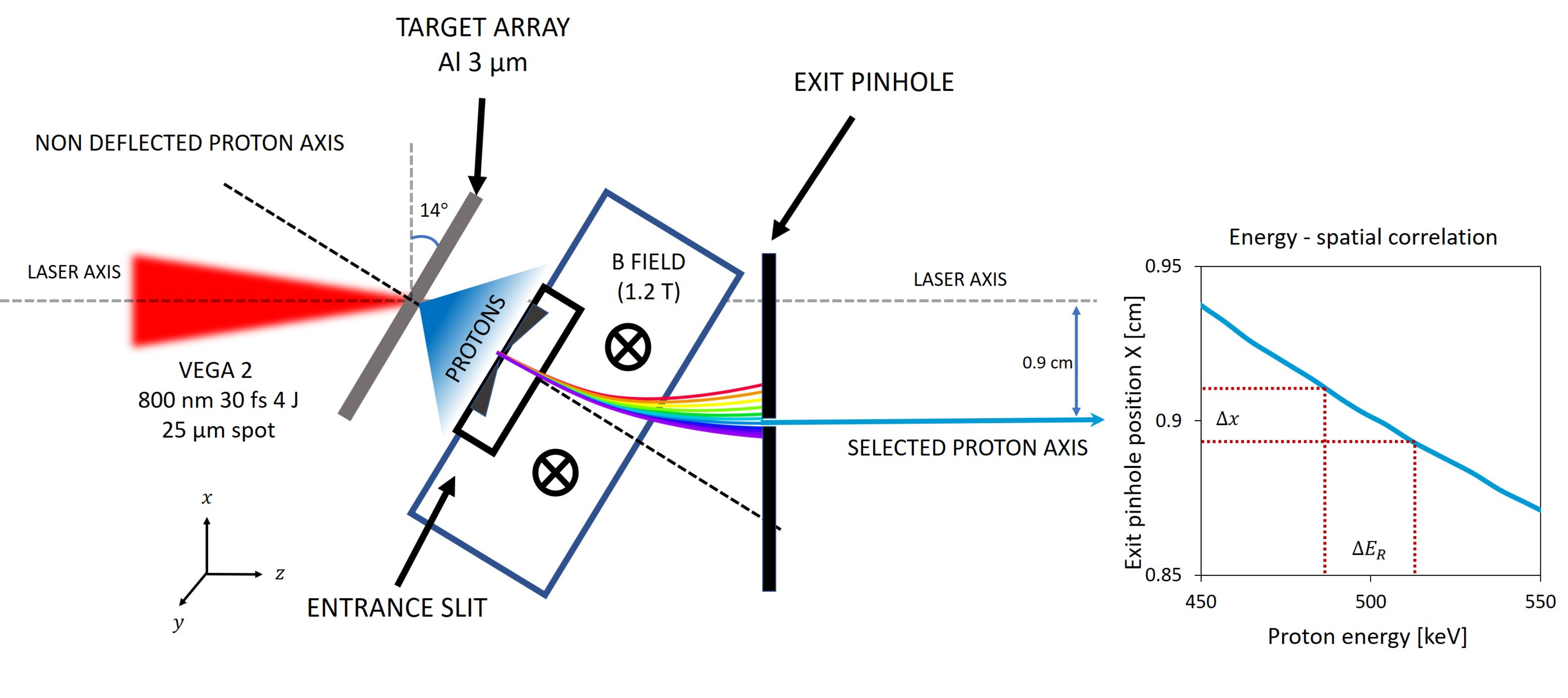}
\caption{Scheme of the energy selection of 0.5 MeV proton beam with the energy spatial correlation.}
\label{fig:Setup_selector}
\end{figure}

The proton selector shown in Figure \ref{fig:Setup_selector} was designed to work with proton energies $<$ 2 MeV and in particular as an adjustable platform for proton stopping power measurements. The system is designed in order to optimize the energy bandwidth and the time spread of the selected beam as a function of the initial beam divergence and spectrum. 
The proton energy selection is based on a magnetic selector working scheme with a dipole for horizontal deflection and two apertures - one before the magnet (entrance slit) and other at the exit of the magnet (exit slit or pinhole).

The entrance slit is directly attached to the dipole magnet yoke, located 1.6 cm from the proton source. It reduces the horizontal acceptance of the laser-driven TNSA proton beam which is characterised by a cone with an opening half-angle around 20$^{\circ}$ and selects a narrow pencil-like proton beam to enter the magnetic field zone. The rectangular slit is exchangeable to allow selection of different apertures of 20$\mu$m and 50 $\mu$m in horizontal and 3 mm vertical. The slits are from Thorlabs and the material is 50  $\mu$m  thick stainless steel 300. This selected beam then enters the field zone and is  dispersed in the horizontal plane according to its energy. 

The exit pinhole is located 1 cm after the exit of the magnet for the selection of a narrow bandwidth of proton energies of 500 keV as shown in the Figure \ref{fig:Exp_MCPSpectrum} b). Other energies can be selected by sliding the pinhole holder horizontally. The set of 10, 20, 200 $\mu$m diameter pinholes is mounted in a motorized holder for fast change between pinholes of different diameters. Using a pinhole as the second filter instead of another slit allows vertical control of the selected beam tilt angle. For these reasons the  exit pinhole is motorized in the horizontal direction (to allow change of the selected energy) and in the vertical direction (to allow tilt control and pinhole exchange). 

\begin{figure}[ht]
\centering
\includegraphics[width=0.9\linewidth]{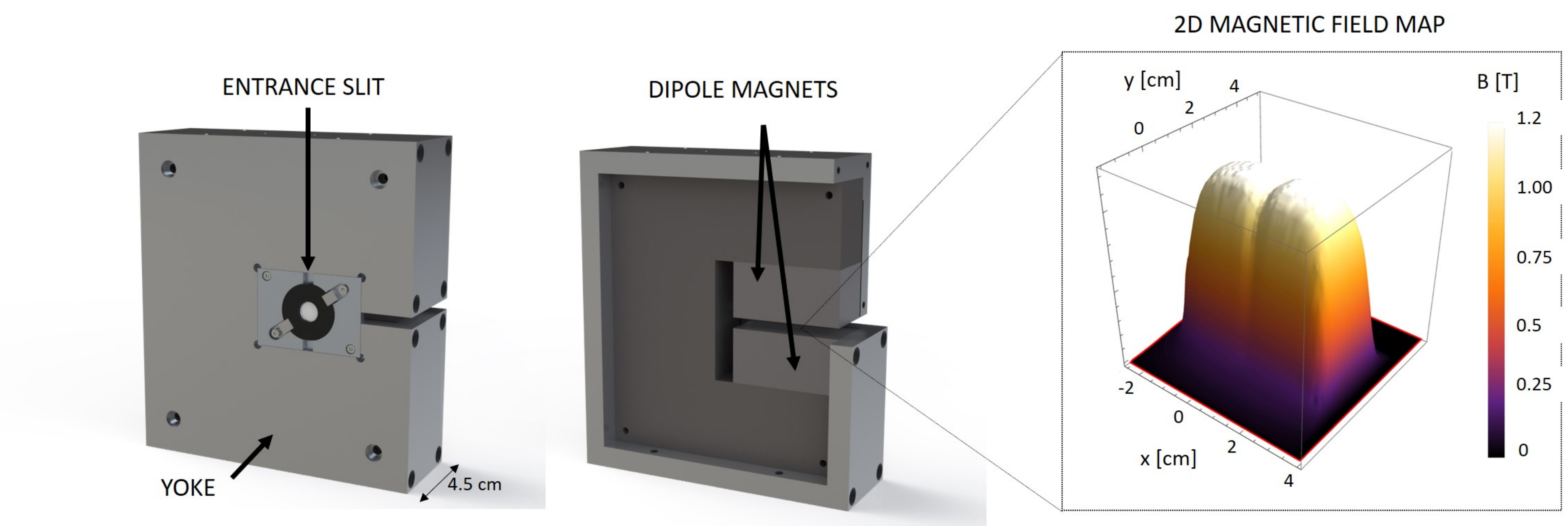}
\caption{The design of magnet with entrance slit for the energy selector. 2D Plot of the magnetic field generated by dipole magnet.}
\label{fig:Setup_magnet}
\end{figure}

The magnet for the energy selector was designed as a part of an internal project of CLPU on the development of high repetition rate diagnostics and secondary sources. It consists of a magnetic alloy yoke and two sets of 2.5 cm permanent magnets separated by a 5 mm gap in order to provide constant magnetic field across a volume of 50.8 $\times$ 25.4 $\times$ 5 mm$^3$, as shown in Figure \ref{fig:Setup_magnet}. The 2D magnetic field map is also shown in Figure  \ref{fig:Setup_magnet}. It was measured with a Hall Effect Gaussmeter probe (LakeshoreTM) with a 1.02 mm active area over a range of 55 mm. Since protons in the 0.1 to 1 MeV range are easily deflected, step-like geometry fields (high squareness) are important to avoid proton deflection at long distances before and after the gap. The selector magnet is motorized in the transverse direction to provide in-out movement for the acquisition of full TNSA spectra for reference purposes when it was out. This motor axis was used as well for precise adjustment of magnet position during the runs, and the vertical adjustment was manual. 

The selected proton beam energy bandwidth is determined by the entrance slit and exit pinhole. Smaller apertures and bigger separations between entrance slit and exit pinhole can provide smaller bandwidths but also increase the time spread and decrease the number of protons. Therefore one should consider a  trade-off between desired low bandwidth and small time spread and minimum required proton flux for application purposes.
The proton pulse, when generated, has a few ps time duration 
\cite{Dromey:2016} but then it acquires time dispersion due to the different velocities of the protons in the bandwidth. Such time spread was limited  to few hundreds of ps by reducing the total selector length within 7-8 cm.
The designed energy selector is compact with a length of 6.2 cm along the proton propagation axis and it can be located as close as needed to the proton source. Another important aspect affecting the ion selection, transport and measurement is due to the initial intrinsic divergence of the accelerated beam. As previously described, such divergence is controlled by using two pinholes one at the entrance and one at the selector exit respectively.

\subsection*{Spectrometer}

\begin{figure}[ht]
\centering
\includegraphics[width=0.6\linewidth]{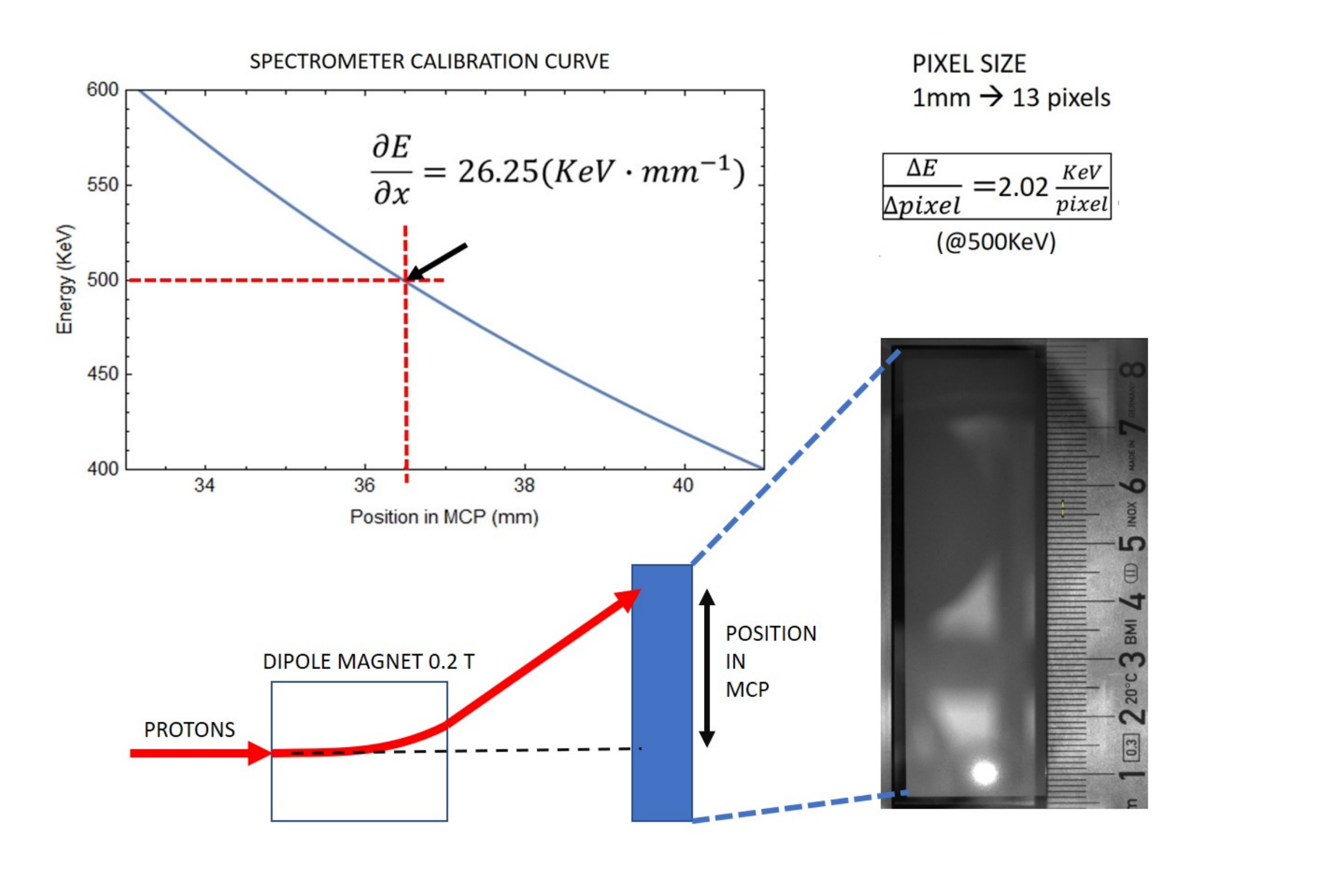}
\caption{Sketch of the spectrometer. Simple expression of dispersion is given as a guide for the reader using a constant magnetic field gyroradius method.}
\label{fig:Setup_spectrometer}
\end{figure}

A magnetic spectrometer was used to measure the central energy, the bandwidth and the profile of the proton beam. Following the scheme in Figure \ref{fig:Setup_spectrometer}, where the deflection as a function of magnetic field and geometry is given.

The spectrometer was positioned 39 cm away from the exit pinhole of the proton energy selector along the propagation axis (Figure \ref{fig:Setup}). It consisted of a magnetic dipole generating a 0.2 T horizontal magnetic field 10.4 cm in length, and a detector composed of a Multichannel-Plate (MCP) coupled with a phosphor screen located 10 cm away from the end of magnet and imaged onto a CCD camera. 
The spectrometer magnet yoke and MCP were enclosed in a lead and Teflon shielding structure to protect them from hard X-ray background. The clear aperture of the spectrometer is defined by a 1 cm diameter entrance hole. A horizontal slit of 1 mm aperture was located outside the shielding just in front of the entrance aperture and ensured that only protons in the horizontal plane of the experiment enter to the spectrometer, i.e. to set the "zero height" (zero deflection) position of the spectrometer. The horizontal slit was motorized in and out in order to locate the proton beam on the detector in case of misalignment, and the height of aperture was adjustable manually. This "zero" position was located in the MCP image using the shadowgraphic X-ray image of the slit observed from the proton source plasma with the magnet removed, 

Figure \ref{fig:Setup_spectrometer} shows the spectrometer calibration curve of energy versus position in the range of interest. The magnetic field map was scanned with the same Hall effect probe used for the selector magnet mapping. With the geometry and the fields of the spectrometer the energy resolution at 500 keV is 2.02 keV per pixel of the image.

\subsection*{Numerical simulations}

In the example simulation setup 6$\times$ $10^{8}$  primary particles sources were used  with an initial TNSA-like spectrum with a maximum energy of 2.5 MeV, an homogeneous angular distribution spread over 30 mrad and a circular source diameter of 150 $\mu$m. The simulation box corresponded to the experimental geometry of the energy selector and magnet spectrometer. A uniform selector magnetic field of $B_{y}$= 1.2 T was used in the simulation while the magnetic field of spectrometer is $B_{x}$=0.2 T.

\bibliography{main_Malko}

\begin{thebibliography}{10}
\urlstyle{rm}
\expandafter\ifx\csname url\endcsname\relax
  \def\url#1{\texttt{#1}}\fi
\expandafter\ifx\csname urlprefix\endcsname\relax\def\urlprefix{URL }\fi
\expandafter\ifx\csname doiprefix\endcsname\relax\def\doiprefix{DOI: }\fi
\providecommand{\bibinfo}[2]{#2}
\providecommand{\eprint}[2][]{\url{#2}}

\bibitem{Neely:2015}
\bibinfo{author}{Danson, C.}, \bibinfo{author}{Hillier, N., D.and~Hopps} \&
  \bibinfo{author}{Neely, D.}
\newblock \bibinfo{journal}{\bibinfo{title}{Petawatt class lasers worldwide}}.
\newblock {\emph{\JournalTitle{High Power Laser Science and Engineering}}}
  \textbf{\bibinfo{volume}{3}}, \doiprefix\url{doi:10.1017/hpl.2014.52}
  (\bibinfo{year}{2015}).

\bibitem{Wilks:2001}
\bibinfo{author}{Wilks, S.~C.} \emph{et~al.}
\newblock \bibinfo{journal}{\bibinfo{title}{Energetic proton generation in
  ultra-intense laser–solid interactions}}.
\newblock {\emph{\JournalTitle{Physics of Plasmas}}}
  \textbf{\bibinfo{volume}{8}}, \bibinfo{pages}{542},
  \doiprefix\url{https://doi.org/10.1063/1.1333697} (\bibinfo{year}{2001}).

\bibitem{Macchi:2013}
\bibinfo{author}{Macchi, A.}, \bibinfo{author}{Borghesi, M.} \&
  \bibinfo{author}{Passoni, M.}
\newblock \bibinfo{journal}{\bibinfo{title}{Ion acceleration by superintense
  laser-plasma interaction}}.
\newblock {\emph{\JournalTitle{Review of Modern Physics}}}
  \textbf{\bibinfo{volume}{85}}, \bibinfo{pages}{42} (\bibinfo{year}{2013}).

\bibitem{Higginson:2018}
\bibinfo{author}{Higginson, A.} \emph{et~al.}
\newblock \bibinfo{journal}{\bibinfo{title}{Near-100 mev protons via a
  laser-driven transparency-enhanced hybrid acceleration scheme}}.
\newblock {\emph{\JournalTitle{Nature Communications}}}
  \textbf{\bibinfo{volume}{9}},
  \doiprefix\url{https://doi.org/10.1038/s41467-018-03063-9}
  (\bibinfo{year}{2018}).

\bibitem{Ken:2014}
\bibinfo{author}{Ledhingham, K.}, \bibinfo{author}{Bolton, P.},
  \bibinfo{author}{Shikazono, N.} \& \bibinfo{author}{Charlie~Ma, C.}
\newblock \bibinfo{journal}{\bibinfo{title}{Towards laser driven hadron cancer
  radiotherapy: A review of progress}}.
\newblock {\emph{\JournalTitle{Progress. Appl. Sci.}}}
  \textbf{\bibinfo{volume}{4}},
  \doiprefix\url{https://doi.org/10.3390/app4030402} (\bibinfo{year}{2014}).

\bibitem{Malka:2019}
\bibinfo{author}{Bayart, E.} \emph{et~al.}
\newblock \bibinfo{journal}{\bibinfo{title}{Fast dose fractionation using
  ultra-short laser accelerated proton pulses can increase cancer cell
  mortality, which relies on functional parp1 protein}}.
\newblock {\emph{\JournalTitle{Scientific Reports}}}
  \textbf{\bibinfo{volume}{9}},
  \doiprefix\url{https://doi.org/10.1038/s41598-019-46512-1}
  (\bibinfo{year}{2019}).

\bibitem{Patel:2013}
\bibinfo{author}{Patel, P.~K.} \emph{et~al.}
\newblock \bibinfo{journal}{\bibinfo{title}{Isochoric heating of solid-density
  matter with an ultrafast proton beam}}.
\newblock {\emph{\JournalTitle{Physical Review Letters}}}
  \textbf{\bibinfo{volume}{91}}, \doiprefix\url{10.1103/PhysRevLett.91.125004}
  (\bibinfo{year}{2003}).

\bibitem{McGuffey:2020}
\bibinfo{author}{McGuffey, C.}, \bibinfo{author}{Kim, J.},
  \bibinfo{author}{Wei, M.} \& \bibinfo{author}{et~al.}
\newblock \bibinfo{journal}{\bibinfo{title}{Focusing protons from a kilojoule
  laser for intense beam heating using proximal target structures.}}
\newblock {\emph{\JournalTitle{Sci Rep}}} \textbf{\bibinfo{volume}{10}},
  \doiprefix\url{10.1038/s41598-020-65554-4} (\bibinfo{year}{2020}).

\bibitem{Pelka:2010}
\bibinfo{author}{Pelka, A.} \emph{et~al.}
\newblock \bibinfo{journal}{\bibinfo{title}{Ultrafast melting of carbon induced
  by intense proton beams}}.
\newblock {\emph{\JournalTitle{Physical Review Letters}}}
  \textbf{\bibinfo{volume}{105}}, \bibinfo{pages}{265701},
  \doiprefix\url{10.1103/PhysRevLett.105.265701} (\bibinfo{year}{2010}).

\bibitem{Bhutwala:2020}
\bibinfo{author}{Bhutwala, K.}, \bibinfo{author}{Bailly-GrandVaux, M.} \&
  \bibinfo{author}{Joohwan, e.~a.}
\newblock \bibinfo{journal}{\bibinfo{title}{Development of a platform at the
  matter in extreme conditions end station for characterization of matter
  heated by intense laser-accelerated protons}}.
\newblock {\emph{\JournalTitle{Transactions on Plasma Science}}}
  \textbf{\bibinfo{volume}{48}}, \bibinfo{pages}{2751--2758},
  \doiprefix\url{10.1109/TPS.2020.3009639.} (\bibinfo{year}{2020}).

\bibitem{Mackinnon:2006}
\bibinfo{author}{Mackinnon, A.} \emph{et~al.}
\newblock \bibinfo{journal}{\bibinfo{title}{Proton radiography of a
  laser-driven implosion}}.
\newblock {\emph{\JournalTitle{Physical Review Letters}}}
  \textbf{\bibinfo{volume}{97}}, \doiprefix\url{10.1103/PhysRevLett.97.045001}
  (\bibinfo{year}{2006}).

\bibitem{Volpe:2011}
\bibinfo{author}{Volpe, L.} \emph{et~al.}
\newblock \bibinfo{journal}{\bibinfo{title}{Proton radiography of laser-driven
  imploding target in cylindrical geometry}}.
\newblock {\emph{\JournalTitle{Physics of Plasmas}}}
  \textbf{\bibinfo{volume}{18}},
  \doiprefix\url{https://doi.org/10.1063/1.3530596} (\bibinfo{year}{2011}).

\bibitem{Hurricane:2016}
\bibinfo{author}{Hurricane, O.} \& \bibinfo{author}{Callahan, D. e.~a.}
\newblock \bibinfo{journal}{\bibinfo{title}{Inertially confined fusion plasmas
  dominated by alpha-particle self-heating}}.
\newblock {\emph{\JournalTitle{Nature Physics}}} \textbf{\bibinfo{volume}{12}},
  \bibinfo{pages}{800--806} (\bibinfo{year}{2016}).

\bibitem{Zylastra:2019}
\bibinfo{author}{Zylstra, A.~B.} \& \bibinfo{author}{Hurricane, O.}
\newblock \bibinfo{journal}{\bibinfo{title}{On alpha-particle transport in
  inertial fusion}}.
\newblock {\emph{\JournalTitle{Physics of Plasmas}}}
  \textbf{\bibinfo{volume}{26}}, \bibinfo{pages}{062701}, \doiprefix\url{doi:
  10.1063/1.5101074} (\bibinfo{year}{2019}).

\bibitem{Roth:2001}
\bibinfo{author}{Roth, M.} \emph{et~al.}
\newblock \bibinfo{journal}{\bibinfo{title}{Fast ignition by intense
  laser-accelerated proton beam}}.
\newblock {\emph{\JournalTitle{Physical Review Letters}}}
  \textbf{\bibinfo{volume}{86}}, \doiprefix\url{10.1103/PhysRevLett.86.436}
  (\bibinfo{year}{2001}).

\bibitem{Fernandez:2014}
\bibinfo{author}{Fernadez, J.} \emph{et~al.}
\newblock \bibinfo{journal}{\bibinfo{title}{Fast ignition with laser-driven
  proton and ion beams}}.
\newblock {\emph{\JournalTitle{Nuclear Fusion}}} \textbf{\bibinfo{volume}{54}}
  (\bibinfo{year}{2014}).

\bibitem{Hoffman:2018}
\bibinfo{author}{Hoffmann, I.}
\newblock \bibinfo{journal}{\bibinfo{title}{Review of accelerator driven heavy
  ion nuclear fusion}}.
\newblock {\emph{\JournalTitle{Matter and Radiation Extremes}}}
  \textbf{\bibinfo{volume}{3}} (\bibinfo{year}{2018}).

\bibitem{Toncian:2006}
\bibinfo{author}{Toncian, T.}, \bibinfo{author}{Borghessi, M.} \&
  \bibinfo{author}{Fuchs, F. e.~a.}
\newblock \bibinfo{journal}{\bibinfo{title}{Ultrafast laser-driven microlens to
  focus and energy-select mega-electron volt protons}}.
\newblock {\emph{\JournalTitle{Science}}} \textbf{\bibinfo{volume}{312}},
  \bibinfo{pages}{410--413}, \doiprefix\url{DOI: 10.1126/science.1124412}
  (\bibinfo{year}{2006}).

\bibitem{Chen:2014}
\bibinfo{author}{Chen, S.} \emph{et~al.}
\newblock \bibinfo{journal}{\bibinfo{title}{Monochromatic short pulse laser
  produced ion beam using a compact passive magnetic device}}.
\newblock {\emph{\JournalTitle{Review of Scientific Instruments}}}
  \textbf{\bibinfo{volume}{85}},
  \doiprefix\url{https://doi.org/10.1063/1.4870250} (\bibinfo{year}{2014}).

\bibitem{Teng:2016}
\bibinfo{author}{Teng, J.} \emph{et~al.}
\newblock \bibinfo{journal}{\bibinfo{title}{Magnetic quadrupoles lens for hot
  spot proton imaging in inertial confinement fusion}}.
\newblock {\emph{\JournalTitle{Nuclear Instruments and Methods in Physics
  Research Section A: Accelerators, Spectrometers, Detectors and Associated
  Equipment}}} \textbf{\bibinfo{volume}{826}}, \bibinfo{pages}{15 -- 20},
  \doiprefix\url{https://doi.org/10.1016/j.nima.2016.03.114}
  (\bibinfo{year}{2016}).

\bibitem{Jahn:2019}
\bibinfo{author}{Jahn, D.} \emph{et~al.}
\newblock \bibinfo{journal}{\bibinfo{title}{Focusing of multi-mev,
  subnanosecond proton bunches from a laser-driven source}}.
\newblock {\emph{\JournalTitle{Phys. Rev. Accel. Beams}}}
  \textbf{\bibinfo{volume}{22}}, \bibinfo{pages}{011301},
  \doiprefix\url{10.1103/PhysRevAccelBeams.22.011301} (\bibinfo{year}{2019}).

\bibitem{Brack:2020}
\bibinfo{author}{Brack, F.}, \bibinfo{author}{Kroll, F.},
  \bibinfo{author}{Gaus, L.} \& \bibinfo{author}{et~al.}
\newblock \bibinfo{journal}{\bibinfo{title}{Focusing of multi-mev,
  subnanosecond proton bunches from a laser-driven source}}.
\newblock {\emph{\JournalTitle{Sci Rep}}} \textbf{\bibinfo{volume}{10}},
  \bibinfo{pages}{13403} (\bibinfo{year}{2020}).

\bibitem{Volpe:2019}
\bibinfo{author}{Volpe, L.} \emph{et~al.}
\newblock \bibinfo{journal}{\bibinfo{title}{Generation of high energy
  laser-driven electron and proton sources with the 200 tw system vega 2 at the
  centro de laseres pulsados}}.
\newblock {\emph{\JournalTitle{High Power Laser Science And Engineering}}}
  \textbf{\bibinfo{volume}{7}},
  \doiprefix\url{https://doi.org/10.1017/hpl.2019.10} (\bibinfo{year}{2019}).

\bibitem{Cayzac:2017}
\bibinfo{author}{Cayzac, W.} \emph{et~al.}
\newblock \bibinfo{journal}{\bibinfo{title}{Experimental discrimination of ion
  stopping models near the bragg peak in highly ionized matter}}.
\newblock {\emph{\JournalTitle{Nature Communications}}}
  \textbf{\bibinfo{volume}{8}}, \bibinfo{pages}{15693},
  \doiprefix\url{https://doi.org/10.1038/ncomms15693} (\bibinfo{year}{2017}).

\bibitem{Frenje:2019}
\bibinfo{author}{Frenje, J.} \emph{et~al.}
\newblock \bibinfo{journal}{\bibinfo{title}{Experimental validation of low-z
  ion-stopping formalisms around the bragg peak in high-energy-density
  plasmas}}.
\newblock {\emph{\JournalTitle{Physical Review Letters}}}
  \textbf{\bibinfo{volume}{122}},
  \doiprefix\url{10.1103/PhysRevLett.122.015002} (\bibinfo{year}{2019}).

\bibitem{Zylstra:2015}
\bibinfo{author}{Zylstra, A.} \emph{et~al.}
\newblock \bibinfo{journal}{\bibinfo{title}{Measurement of charged-particle
  stopping in warm dense plasma}}.
\newblock {\emph{\JournalTitle{Physical Review Letters}}}
  \textbf{\bibinfo{volume}{114}},
  \doiprefix\url{10.1103/PhysRevLett.114.215002} (\bibinfo{year}{2015}).

\bibitem{Malko:2020}
\bibinfo{author}{Malko, S.}
\newblock \bibinfo{journal}{\bibinfo{title}{Charged particle transport in
  extreme state of matter}}.
\newblock {\emph{\JournalTitle{PhD Thesis, University of Salamanca}}}
  (\bibinfo{year}{2020}).

\bibitem{Prasad:2013}
\bibinfo{author}{Prasad, R.} \emph{et~al.}
\newblock \bibinfo{journal}{\bibinfo{title}{Thomson spectrometer–microchannel
  plate assembly calibration for mev-range positive and negative ions, and
  neutral atoms}}.
\newblock {\emph{\JournalTitle{Rev. Sci. Instrum.}}}
  \textbf{\bibinfo{volume}{84}}, \doiprefix\url{10.1063/1.4803670}
  (\bibinfo{year}{2013}).

\bibitem{Rossi:1941}
\bibinfo{author}{Rossi, B.} \& \bibinfo{author}{Greisen, K.}
\newblock \bibinfo{journal}{\bibinfo{title}{Cosmic ray theory}}.
\newblock {\emph{\JournalTitle{Reviews of Modern Physics}}}
  \textbf{\bibinfo{volume}{13}}, \bibinfo{pages}{240--309}
  (\bibinfo{year}{1941}).

\bibitem{SRIM}
\bibinfo{author}{Ziegler, J.~F.}, \bibinfo{author}{Ziegler, M.~D.} \&
  \bibinfo{author}{Biersack, J.~P.}
\newblock \bibinfo{journal}{\bibinfo{title}{Srim – the stopping and range of
  ions in matter (2010)}}.
\newblock {\emph{\JournalTitle{Nuclear Instruments and Methods in Physics
  Research B}}} \textbf{\bibinfo{volume}{268}}, \bibinfo{pages}{1818--1823},
  \doiprefix\url{10.1016/j.nimb.2010.02.091} (\bibinfo{year}{2010}).

\bibitem{Bohlen:2014}
\bibinfo{author}{Böhlen, T.} \emph{et~al.}
\newblock \bibinfo{journal}{\bibinfo{title}{The fluka code: Developments and
  challenges for high energy and medical applications}}.
\newblock {\emph{\JournalTitle{Nuclear Data Sheets}}}
  \textbf{\bibinfo{volume}{120}} (\bibinfo{year}{2014}).

\bibitem{Ferrari:2005}
\bibinfo{author}{Ferrari, A.}, \bibinfo{author}{Sala, P.},
  \bibinfo{author}{Fasso, A.} \& \bibinfo{author}{Ranft, J.}
\newblock \bibinfo{journal}{\bibinfo{title}{Fluka: a multi-particle transport
  code}}.
\newblock {\emph{\JournalTitle{CERN-2005-10}}}  (\bibinfo{year}{2005}).

\bibitem{Borghesi:2004}
\bibinfo{author}{Borghesi, M.} \emph{et~al.}
\newblock \bibinfo{journal}{\bibinfo{title}{Multi-mev proton source
  investigations in ultraintense laser-foil interactions}}.
\newblock {\emph{\JournalTitle{Phys. Rev. Lett.}}}
  \textbf{\bibinfo{volume}{92}}, \bibinfo{pages}{055003},
  \doiprefix\url{10.1103/PhysRevLett.92.055003} (\bibinfo{year}{2004}).

\bibitem{Borghesi:2006}
\bibinfo{author}{Borghesi, M.} \emph{et~al.}
\newblock \bibinfo{journal}{\bibinfo{title}{Fast ion generation by
  high-intensity laser irradiation of solid targets and applications}}.
\newblock {\emph{\JournalTitle{Fusion Science and Technology}}}
  \textbf{\bibinfo{volume}{49}}, \bibinfo{pages}{412--439},
  \doiprefix\url{10.13182/FST06-A1159} (\bibinfo{year}{2006}).

\bibitem{Covan:2004}
\bibinfo{author}{Cowan, T.~E.} \emph{et~al.}
\newblock \bibinfo{journal}{\bibinfo{title}{Ultralow emittance, multi-mev
  proton beams from a laser virtual-cathode plasma accelerator}}.
\newblock {\emph{\JournalTitle{Phys. Rev. Lett.}}}
  \textbf{\bibinfo{volume}{92}}, \bibinfo{pages}{204801},
  \doiprefix\url{10.1103/PhysRevLett.92.204801} (\bibinfo{year}{2004}).

\bibitem{Malko:2019}
\bibinfo{author}{Malko, S.}, \bibinfo{author}{Vaisseau, X.},
  \bibinfo{author}{Perez, F.} \& \bibinfo{author}{et~al.}
\newblock \bibinfo{journal}{\bibinfo{title}{Enhanced relativistic-electron beam
  collimation using two consecutive laser pulses}}.
\newblock {\emph{\JournalTitle{Sci Rep}}} \textbf{\bibinfo{volume}{9}},
  \bibinfo{pages}{14061} (\bibinfo{year}{2019}).

\bibitem{Buffechoux:2010}
\bibinfo{author}{Buffechoux, S.} \emph{et~al.}
\newblock \bibinfo{journal}{\bibinfo{title}{Hot electrons transverse refluxing
  in ultraintense laser-solid interactions}}.
\newblock {\emph{\JournalTitle{Phys. Rev. Lett.}}}
  \textbf{\bibinfo{volume}{105}}, \bibinfo{pages}{015005},
  \doiprefix\url{10.1103/PhysRevLett.105.015005} (\bibinfo{year}{2010}).

\bibitem{Dromey:2016}
\bibinfo{author}{Dromey, B.} \emph{et~al.}
\newblock \bibinfo{journal}{\bibinfo{title}{Picosecond metrology of
  laser-driven proton bursts}}.
\newblock {\emph{\JournalTitle{Nature Communications}}}
  \textbf{\bibinfo{volume}{7}},
  \doiprefix\url{https://doi.org/10.1038/ncomms10642} (\bibinfo{year}{2016}).

\end{thebibliography}

\section*{Acknowledgements}

The research leading to these results has received funding from LASERLAB-EUROPE (grant agreement no. 654148, European UnionAos Horizon 2020 research and innovation program). Also, we would like to thank the Ministry of Science, the Castilla y Leon region and the University of Salamanca for the constant support during all the phases of construction and implementation of the CLPU facility. Support from Spanish Ministerio de Ciencia, Innovación y Universidades through the PALMA Grant No.
FIS2016-81056-R, ICTS Equipment Grant No. EQC2018-005230-P; from LaserLab Europe IV Grant No. 654148 and from Junta de Castilla y León Grant No. CLP087U16 is acknowledged.

\section*{Author contributions statement}

L. Volpe and S. Malko conceived the idea and led the experiment. S. Malko, J. Apiñaniz, R. Fedosejevs, W. Cayzac, X. Vaisseau and L. Volpe worked on the experimental proposal, the experimental design, the initial design of the magnetic selector and on the analysis of the experimental results. The ultimate design and implementation of the selector was done by G. Gatti and D. de Luis. S. Malko and L. Volpe performed the Monte Carlo simulations for the interpretation of the results. 
J.J. Apiñaniz, S. Malko,R. Fedosejevs, J. A. Perez and L. Volpe performed the experiment at the CLPU. 
X. Vaisseau, W. Cayzac, M. Bailly-Grandvaux, K. Bhutwala, C. McGuffey, V. Ospina and J. Balboa participated in the analysis of the experimental results. J.J. Santos, D. Batani, F. Beg and L. Roso were supporting the activities with precious discussions and suggestions.  
J. Apiñaniz, S. Malko, R. Fedosejevs and L. Volpe wrote the first draft of the paper. All the authors contributed to the final version of the manuscript. 

\end{document}